\documentclass[aps,preprintnumbers,reprint,showkeys,eqsecnum,nofootinbib,floatfix,amsmath,asmfonts,latexsym,amssymb]{revtex4-1}
\usepackage{hyperref,graphicx,slashed,color,multirow,float,amsfonts,bbold,mathtools}
\pdfoutput=1
\allowdisplaybreaks
\newcommand{\nubarnu}{\raisebox{1ex}{\hbox{\tiny(}}\overline\nu\raisebox{1ex}{\hbox{\tiny)}}\hspace{-0.5ex}}

\usepackage{ulem}
\let\emph\textit

\begin{document}
\preprint{IP/BBSR/2021-0X}
\title{Type-III Seesaw: Phenomenological Implications of the Information Lost in Decoupling from High-Energy to Low-Energy}
\author{Saiyad Ashanujjaman}
\email[\href{https://orcid.org/0000-0001-5643-2652}{\includegraphics[scale=0.4]{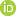}}~~]{saiyad.a@iopb.res.in}
\author{Kirtiman Ghosh}
\email[]{kirti.gh@gmail.com}
\affiliation{Institute of Physics, Bhubaneswar, Sachivalaya Marg, Sainik School, Bhubaneswar 751005, India}%
\affiliation{Homi Bhabha National Institute, Training School Complex, Anushakti Nagar, Mumbai 400085, India}%

\begin{abstract}
\noindent The type-III seesaw seems to explain the very minuteness of neutrino masses readily and naturally. The high-energy see-saw theories usually involve larger number of effective parametres than the physical and measurable parametres appearing in the low-energy neutrino phenomenology. Casas-Ibarra parametrisation facilitates to encode the information lost in integrating the heavy fermions out in an arbitrary complex orthogonal matrix. The CMS collaboration has already searched for triplet fermions in the type-III seesaw model with only one generation of triplet fermion flavour democratically decaying into SM leptons. We reinterpret this CMS search in the context of a realistic type-III seesaw model with two or three generations of triplet fermions, and endeavour to comprehend the implications of the foregoing matrix on the $95\%$ CL lower limit on the mass of the triplet fermions. We also discuss the phenomenological implication of the aforesaid matrix in view of charged lepton flavour violating observables and displaced decays of the triplet fermions at colliders.
\end{abstract}
\maketitle

\section{\label{sec:intro}Introduction}
\noindent Notwithstanding Standard Model(SM)'s mightily success, it falls short in addressing the issue of non-zero tiny neutrino masses and mixing. Although plausible, it seems flimsy that the tiny neutrino masses are generated by dint of the usual Brout, Englert and Higgs mechanism as it entails minute Yukawa couplings which render philosophical and aesthetic displeasure. Diversely, the three UV completions corresponding to the so-called Weinberg operator \cite{seesaw_gut_2} at tree level \cite{seesaw_gut_5} --- type-I \cite{seesaw1_1,seesaw1_2,seesaw1_3,seesaw1_4,seesaw1_5}, type-II \cite{seesaw2_1,seesaw2_2,seesaw2_3,seesaw2_4,seesaw2_5,seesaw2_6} and type-III \cite{seesaw3_1} seesaw seem to accommodate the observed sub-eV neutrino masses more readily and naturally. Though the fields which mediate neutrino masses in classical seesaw are naturally motivated to have very high scale masses, nothing preclude them to have masses at collider testable TeV scale if balanced with a small Yukawa coupling. In this work, we focus on the type-III seesaw model \cite{typeIII_1,typeIII_2,typeIII_3,Aguila,Aguila2,Hambye,TaoHan} which employs TeV scale $SU(2)_L$ triplet chiral fermions with zero hypercharge. 

Given that the lightest neutrino mass is still undetermined, the low-energy effective theory involves 9(7) physical and measurable parametres for non-zero (zero) lightest neutrino mass, whereas the high-energy seesaw theory involves 15(9) effective parametres in 3RHN(2RHN) case\footnote{We refer the scenario when the SM is extended by three(two) generations of right-handed triplet fermions as 3RHN(2RHN) case.}(see Section \ref{sec:model}). A number of parametres get lost in integrating the heavy fields out. Casas-Ibarra parametrisation \cite{Casas_Ibarra,Ibarra_Ross} facilitates to encode these otherwise lost information in an arbitrary complex orthogonal matrix. To paraphrase it, the gap between the low- and high-energy is bridged via an arbitrary complex orthogonal matrix. Mostly, in the literature \cite{Aguila, Aguila2, Hambye, Deepa, Bpriyo}, this matrix has been assumed imprudently to be identity ignoring the possibility of its non-triviality and hence its phenomenological implications. Also, most of the studies, especially the experimental searches \cite{ex0, ex1, ex2, ex3, ex4, ex5, ex6, cms_137}, considered a simplified type-III scenario with only one generation of triplet fermions flavour democratically decaying into leptons, albeit a realistic type-III scenario requires two or more generations of triplet fermions. In this work, we inspect the phenomenological implications of the said matrix in view of lepton flavour violation, displaced decays and a recent CMS search \cite{cms_137} for type-III seesaw triplets at the Large Hadron Collider (LHC) with $\sqrt s=13$ TeV and 137.1 fb$^{-1}$ integrated luminosity.

\section{\label{sec:model} The Type-III Seesaw Model}
\noindent The field content of the SM is extended by at least two generations of right-handed $SU(2)_L$ triplet fermions ($\Sigma_R$) with zero hypercharge
\begin{equation*}
\Sigma_R = \left( \begin{array}{cc} \Sigma^0_R/\sqrt{2} & \Sigma^+_R \\ \Sigma^-_R & -\Sigma^0_R/\sqrt{2} \end{array} \right)~.
\end{equation*}
The relevant triplet Yukawa and mass terms which generates Weinberg operator at tree level are given by
\begin{equation*}
-{\cal L} \supset M_\Sigma^{ij} {\rm Tr}\left( ~\overline{\tilde{\Sigma}^i_R} \Sigma_R^j \right) + \sqrt{2}~ Y_\Sigma^{i\alpha}~\tilde{H}^\dagger \overline{\Sigma_R^i} L^\alpha ~ + \rm{h.c.}~,
\end{equation*}
where $i,j$ and $\alpha$ are the generation indices -- $i$ and $j$ run over 1 and 2(1,2 and 3) for 2RHN(3RHN) case, and $\alpha$ runs over $e,~\mu$ and $\tau$; $M_\Sigma$ is the Majorana mass matrix for $\Sigma_R$, $\tilde{\Sigma}_R$ denotes the charge-conjugation of $\Sigma_R$, {\it i.e.} $\tilde{\Sigma}_R=C(\overline{\Sigma_R})^T$ with $C$ being the charge-conjugation matrix, $Y_\Sigma$ is the Yukawa matrix (in general, complex) with $Y_\Sigma^{1\alpha}, Y_\Sigma^{2\alpha}$ and $Y_\Sigma^{3\alpha}$ being affiliated to first, second and third generation of triplet fermions, $L=\left(\nu_L, \ell_L \right)^T$ is the left-handed SM lepton doublet, and $H$ is the SM Higgs doublet. After the electroweak symmetry breaking (EWSB), the above Lagrangian leads to mixing between the SM leptons and the heavy triplet leptons which, in turn, leads to the following seesaw formula for light neutrinos:
\begin{equation}
\label{eq:nu_mass}
m_\nu \approx -\frac{v^2}{2} Y_\Sigma^T M_\Sigma^{-1} Y_\Sigma~,
\end{equation} 
where $v$ is the vacuum expectation value for the SM Higgs. Further, $m_\nu$ can be diagonalised using the so-called Pontecorvo-Maki-Nakagawa-Sakata (PMNS) matrix, $U$, as $U^T m_\nu U=D_{m_\nu}={\rm diag}(m_1,m_2,m_3)$, where $U$ is usually parametrised by three (real) angles ($\alpha_{12},\alpha_{23}$ and $\alpha_{13}$), one Dirac phase ($\delta$) and two Majorana phases ($\phi$ and $\phi^\prime$) with $U\times \mathrm{diag}(e^{i \phi/2},e^{i \phi^\prime/2},1)~=$
\begin{equation*}
\left[ \begin{array}{ccc}
1 & 0 & 0 \\
0 & c^{23} & s^{23} \\
0 & -s^{23} & c^{23} 
\end{array} \right] 
\left[\begin{array}{ccc}
c^{13} & 0 & s^{13}e^{-i\delta}\\
0 & 1 & 0 \\
-s^{13}e^{i\delta} & 0 & c^{13} 
\end{array} \right]
\left[ \begin{array}{ccc}
c^{12} & s^{12} & 0 \\
-s^{12} & c^{12} & 0 \\
0 & 0 & 1
\end{array} \right],
\end{equation*}
where $c^{ij}\left(s^{ij}\right)=\cos\alpha_{ij}\left(\sin\alpha_{ij}\right)$. We take $\delta, \phi$ and $\phi^\prime$ to be zero for simplicity throughout this work\footnote{We take these phases to be zero simply because either they are poorly measured or not measured at all. However, a non-zero choice for the phases may significantly alter the phenomenology.}. For normal hierarchy (NH), $m_2=\sqrt{\Delta m_{21}^2+m_1^2}$, $m_3=\sqrt{\Delta m_{31}^2+m_1^2}$, and for inverted hierarchy (IH), $m_1=\sqrt{|\Delta m_{32}^2+\Delta m_{21}^2-m_3^2|}$, $m_2=\sqrt{|\Delta m_{32}^2-m_3^2|}$; where $\Delta m_{ij}^2$'s are the experimentally measured mass-squared differences, and $m_1(m_3)$ is the obscured lightest neutrino mass for NH(IH) subject to the the current bound from cosmology $\sum_i m_i < 0.12$ eV \cite{planck}. The best fit values for the neutrino oscillation parametres used in this work are taken from Ref.~\cite{nu_osci_param}.


Note that the low-energy neutrino oscillation data is completely described in terms of 3(2) neutrino masses, 3 mixing angles and 3(2) phases, {\it i.e.} 9(7) physical and measurable parametres, whereas the high-energy seesaw theory involves 15(9) effective parametres in 3RHN(2RHN) case\footnote{In the basis where $M_\Sigma$ is diagonal, $M_\Sigma$ is determined by 3(2) real parametres and $Y_\Sigma$, being a $3\times3(2\times 3)$ complex matrix, contains 18(12) real parametres in 3RHN(2RHN) case. However, 3 redundant phases in $Y_\Sigma$ can be eliminated by a redefinition of $L$, and 3(2) more parametres can be reduced in view of scaling symmetry of Eq.~\eqref{eq:nu_mass}, thus, we end up with only 15(9) effective parametres in 3RHN(2RHN) case.}. In order to ensure the consistency between the high-energy seesaw theory and the low-energy neutrino phenomenology, proper parametrisation of the Yukawa matrix ($Y_\Sigma$) is required. In the basis where $M_\Sigma$ is diagonal, the most general texture of $Y_\Sigma$ resulting into measured values of the neutrino oscillation parametres is given by \cite{Casas_Ibarra,Ibarra_Ross},
\begin{equation}
Y_\Sigma=i\frac{\sqrt{2}}{v} \sqrt{M_\Sigma} R \sqrt{D_{m_\nu}}U^\dagger~, \hspace{1cm}(i=\sqrt{-1})
\label{eq:casas_ibarra}
\end{equation}
where $M_\Sigma={\rm diag}(M_1,M_2,M_3)[(M_1,M_2)]$ in 3RHN[2 RHN] case, and $R$ is an arbitrary complex orthogonal matrix\footnote{Strictly speaking, in 2RHN case, $R$ is not an orthogonal matrix, withal it satisfies $R^T R=\mathbb{1}$.} which encodes the residual freedom in $Y_\Sigma$ once the other parametres are are determined from the low-energy neutrino oscillation data. To comprehend the phenomenological consequences of $R$, we parametrise the same by three(one) complex angles in 3RHN(2RHN) case. For 2RHN case \cite{Ibarra_Ross},
\begin{equation*}
R=\begin{cases} \left( \begin{array}{ccc}
0 & \cos \theta & \zeta \sin \theta \\
0 & -\sin \theta & \zeta \cos \theta
\end{array} \right) \mathrm{for~NH}~,
\\
\\
\left( \begin{array}{ccc}
\cos \theta & \zeta \sin \theta & 0\\
-\sin \theta & \zeta \cos \theta & 0
\end{array} \right) \mathrm{for~IH}~. \end{cases}
\end{equation*}
where $\zeta = \pm$1\footnote{Throughout this analysis, we take $\zeta=+1$ as $\zeta=-1$ will not yield any physically different texture of  $Y_\Sigma$.} and the complex angle $\theta=x+iy$ with $x,y \in \mathbb{R}$.  
Note that $|(Y_\Sigma)_{ij}|$'s are invariant separately under ${\it (i)}x \to x + \pi$ and ${\it (ii)}y \to -y$\footnote{To be precise, $|(Y_\Sigma)_{ij}|$'s are invariant under $y \to -y$ and $\delta \to -\delta$, but, as mentioned earlier, we take $\delta$ to be zero for simplicity.}. Ergo, it suffices to restrict to $x \in [0,\pi] ~\mathrm{and}~ y \geq 0$.
For 3RHN case \cite{Casas_Ibarra},
\begin{equation*}
R =\left( \begin{array}{ccc}
1 & 0 & 0 \\
0 & c_{23} & s_{23} \\
0 & -s_{23} & c_{23} 
\end{array} \right) 
\left( \begin{array}{ccc}
c_{13} & 0 & s_{13}\\
0 & 1 & 0 \\
-s_{13} & 0 & c_{13} 
\end{array} \right)
\left( \begin{array}{ccc}
c_{12} & s_{12} & 0 \\
-s_{12} & c_{12} & 0 \\
0 & 0 & 1
\end{array} \right)~,
\end{equation*}
where $c_{ij}\left(s_{ij}\right)=\cos \theta_{ij}\left(\sin \theta_{ij}\right)$\footnote{Note that this parametrisation of $R$ is not the most general one as it does not include reflections; however, the uncounted forms of $R$ do not produce any physically different textures of $Y_\Sigma$.} and three complex angles $\theta_{23}=x_1+iy_1$, $\theta_{13}=x_2+iy_2$ and $\theta_{12}=x_3+iy_3$ with $x_{1,2,3},y_{1,2,3} \in \mathbb{R}$. Because of large number of parametres in this case, the effect of $R$ on the Yukawa matrix $Y_\Sigma$ is quite intricate unlike 2RHN case, however, it can be reckoned that $x$'s lead to oscillatory nature of $Y_\Sigma$ and $y$'s enhance $Y_\Sigma$ exponentially. Note that $|(Y_\Sigma)_{ij}|$'s are invariant under $x_1 \to x_1 + \pi$. Therefore, it suffices to restrict to $x_1 \in [0,\pi]$ and $x_{2,3} \in [0,2\pi]$ while keeping $y_{1,2,3}$ unbounded\footnote{Ref.~\cite{Arindam_Sanjoy} take $y_{1,2,3} \in [-\pi,\pi]$ in their analysis. We accentuate that, in general, $y_{1,2,3}$ are unbounded; there is no reason for $y$'s to be bounded in the range $[-\pi,\pi]$.}.

Note that $R_{1j} \to R_{2j}$ ($j=1,2,3$) as $\theta \to \theta + \pi/2$ for 2RHN case, whereas $R_{2j} \to R_{3j}$ ($j=1,2,3$) as $\theta_{23} \to \theta_{23} + \pi/2$ for 3RHN case. This is a consequence of the orthogonality of the $R$ matrix. In fact, this in turn has a phenomenological consequence as we will see later.

\section{\label{sec:lfv}Lepton flavour violation}
\noindent The texture of the triplet Yukawa couplings (see Eq.~\eqref{eq:casas_ibarra}) necessary to generate flavour mixing in the neutrino sector, also gives rise to SM charged lepton flavour violating processes (CLFVPs) like $\mu \to e\gamma$ and $\tau \to \ell \gamma$ ($\ell=e,\mu$) at one-loop level, $\mu \to 3e$ and $\tau \to 3\ell$ at tree level, {\it etc}. The constraints on the matrix $R$ resulting from the experimental limits on the CLFVPs are discussed in the following.

For convenience, it is customary to define $|\epsilon_{\alpha \beta}| = \frac{v^2}{2}|Y_\Sigma^\dagger M_\Sigma^{-2} Y_\Sigma|_{\beta \alpha}$. Among all the CLFVPs, the charge-changing $\mu^- \to e^+$ conversion rate to total nucleon muon capture rate ratio on $^{48}_{22} \rm Ti$ for the ground state transition from SINDRUM-II collaboration at PSI \cite{mueti} puts the most stringent bound\footnote{Note that this limit is better than what is quoted in Ref.~\cite{Biggio_et_al}. Somehow they missed the most improved limit from SINDRUM-II collaboration \cite{mueti}. Also, note the relation between the notations used in Ref.~\cite{Biggio_et_al} and the present work: $|\epsilon_{e\mu}|=2|\eta_{\mu e}|$.} on $|\epsilon_{e\mu}| < 3.7 \times 10^{-7}$ at $2\sigma$. Diversely, $\tau$ decays into three leptons --- $e\mu\mu$ and $\mu ee$ \cite{belle} set the most constraining bounds on $|\epsilon_{e\tau}| < 6.0\times 10^{-4}$ and $|\epsilon_{\mu \tau}| < 5.0\times 10^{-4}$ \cite{Biggio_et_al}, respectively. Because of many free parametres in this scenario, the expressions for $\epsilon_{\alpha \beta}$ become perplexing; hence, we avert to write them down explicitly. Instead, we perform a scanning over the parametres of $R$, namely the $x$'s and $y$'s, and put constraints on them using the aforecited upper bounds on $|\epsilon_{\alpha \beta}|$. We find that, in 2RHN case, for $M_\Sigma \sim$ TeV, $y \lesssim \mathcal{O}(9)$ are allowed for all values of $x$\footnote{This bound is consistent with Ref.~\cite{srubabati_2019}.}.
Whereas in 3RHN case, for $M_\Sigma \sim$ TeV, we find the following constraints:
\begin{enumerate}
\item $y_{1,2,3} \lesssim \mathcal{O}(11)$ and $y_1+y_2+y_3 \lesssim \mathcal{O}(14)$ for all values of $x_{1,2,3}$ except for $x_2 \sim 3\pi/2$.
\item   $y_{1,3} \lesssim \mathcal{O}(45)$, $y_2 \lesssim \mathcal{O}(10)$ and $y_1+y_2+y_3 \lesssim \mathcal{O}(90)$ for $x_2 \sim 3\pi/2$.
\end{enumerate}

\noindent These bounds get stronger with decreasing $M_\Sigma$ as well as increasing the lightest neutrino mass as $|\epsilon_{\alpha \beta}| \propto \mathcal{O}(m_\nu/M_\Sigma)$. Also, the bounds on $y$'s are bit more constraining in IH case than in NH one. Note that the bounds from perturbativity as well as electroweak precision data are much liberal than those obtained from CLFVPs.

\section{\label{sec:pro_dec}Production and Decay Modes of Fermion Triplets}
\noindent The TeV scale triplet fermions are copiously pair produced \cite{typeIII_1,typeIII_2,typeIII_3,Aguila,Aguila2,Hambye,TaoHan} at the LHC \cite{typeIII_1,typeIII_2,typeIII_3,Aguila,Aguila2,Hambye,TaoHan} by quark-antiquark annihilation via $s$-channel $\gamma/Z$ and $W^{±}$ exchanges, namely the Drell-Yan processes\footnote{Also, the charged triplet fermions can be produced via $t/u$-channel photon-photon fusion process. 
  However, their production through such process is sub-dominant for the mass range of our interest.}.
After being produced, the triplets undergo two types of decays --- the heavy state transitions due to the radiative mass-splitting\footnote{$\Delta M=M_{\Sigma^+}-M_{\Sigma^0} \sim 166~\rm MeV$ \cite{cirelli} for $M_\Sigma \gg M_{Z,W}$.} between $\Sigma^\pm$ and $\Sigma^0$ and the SM two-body final state decays induced by their mixing with the SM leptons. The corresponding partial decay widths are given by
\begin{align}
&\Gamma(\Sigma^{\pm}_i \to \Sigma^{0}_i \pi^\pm )= \frac{2G_{\rm F}^2 V_{ud}^2 f_\pi^2 (\Delta M)^3}{\pi}
\sqrt{1-\frac{m_\pi^2}{(\Delta M)^2}}~,  \nonumber
\\
&\Gamma(\Sigma^{\pm}_i \to \Sigma^{0}_i \ell^\pm \nubarnu_\ell )= \frac{2G_F^2 (\Delta M)^5}{15\pi^3} \sqrt{1-\frac{m_\ell^2}{(\Delta M)^2}} ~P\left(\frac{m_\ell}{\Delta M} \right)~, \nonumber
\\
& \Gamma(\Sigma^{0(\pm)}_i\to XY^{(\prime)})  = \kappa \frac{M_i}{64\pi} |(Y_\Sigma)_{i\alpha}|^2 f\left(\frac{M_X}{M_i}\right)~,
\label{eq:DW_Sigma_XY}
\end{align}
where $G_F$ is the Fermi coupling constant, $|V_{ud}|\approx 0.974$ \cite{Vud}, $f_\pi \simeq 130$ MeV \cite{fpi} is the pion decay constant, $m_\pi$ and $m_\ell$ are, respectively, the charged pion and lepton ($\ell=e,\mu$) mass, $P(r)=1-\frac{9}{2}r^2-4r^4+\frac{15r^4}{2\sqrt{1-r^2}} {\rm tanh}^{-1} \sqrt{1-r^2}$~, $XY \ni h \nubarnu_\alpha, Z \nubarnu_\alpha, W^\pm \ell^\mp_\alpha$, $XY^\prime \ni h \ell^\pm_\alpha, Z \ell^\pm_\alpha, W^\pm  \nubarnu_\alpha$, $f(r)=(1-r^2)^2(1+\kappa^\prime r^2)$, and $\kappa(\kappa^\prime)=1,1,2(0,2,2)$ for $X=h,Z,W^\pm$, respectively. The total decay width of $\Sigma^{0,\pm}_i$ are given by\footnote{The contribution of the heavy state transitions to the total decay width of $\Sigma^\pm$ turns out to be $\sim 3.5\times 10^{-15}$ GeV.}:
\begin{eqnarray*}
&&\Gamma_{i,\rm T}^0 \approx \frac{M_i}{16\pi} \sum_\alpha |(Y_\Sigma)_{i \alpha}|^2 = \frac{M_i^2}{8\pi v^2} \sum_j m_j \left|R_{ij}\right|^2 {\rm ~and}
\end{eqnarray*}
$\Gamma_{i,\rm T}^\pm \approx \Gamma_{i,\rm T}^0 + 3.5\times 10^{-15} \mathrm{~GeV}$. The relevant expressions for $\Gamma_{i,\rm T}^0$'s in terms of the parametres in the arbitrary orthogonal complex matrix $R$, namely the $x$'s and $y$'s, are presented in the following. For 2RHN case, the total decay widths for the neutral component of the first and second generation triplets are given by,
\begin{eqnarray}
\Gamma_{1,\rm T}^0 &&\approx \frac{M_1^2}{16\pi v^2} \times
\begin{cases}
\left[ m_2 f^+ + m_3 f^- \right]~{\rm~for~NH}~,
\\
\left[ m_1 f^+ + m_2 f^- \right]~{\rm~for~IH}~,
\end{cases}
\label{eq:2rhn_dl}
\\
\Gamma_{2,\rm T}^0 &&\approx \frac{M_2^2}{16\pi v^2} \times
\begin{cases}
\left[ m_2 f^- + m_3 f^+ \right]~{\rm~for~NH}~,
\\
\left[ m_1 f^- + m_2 f^+ \right]~{\rm~for~IH}~;
\end{cases}
\label{eq:2rhn_dl2}
\end{eqnarray}
For 3RHN case,
\begin{eqnarray}
&&\Gamma_{1,\rm T}^0 \approx \frac{M_1^2}{32\pi v^2} \left[ m_1 f^+_2 f^+_3 + m_2 f^+_2 f^-_3 + 2m_3 f^-_2 \right],
\label{eq:3rhn_dl}
\end{eqnarray}
where $f^\pm_{(k)}=\cosh 2y_{(k)} \pm \cos 2x_{(k)}$ with $k=1,2,3$. One can obtain similar expressions for the total decay widths of the neutral component of the second and third generation triplets for 3RHN case.

We follow from the above expressions that $\Gamma_{1,\rm T}^0/M_1^2 \to \Gamma_{2,\rm T}^0/M_2^2$ as $f^\pm \to f^\mp$, {\it i.e.} $\theta \to \theta + \pi/2$ for 2RHN case. A similar relation, {\it i.e.} $\Gamma_{2,\rm T}^0/M_2^2 \to \Gamma_{3,\rm T}^0/M_3^2$ as $\theta_{23} \to \theta_{23} + \pi/2$, holds for 3RHN case. This is an upshot of the fact that, as mentioned at the closing of Section~\ref{sec:model}, $R_{1j} \to R_{2j}$($R_{2j} \to R_{3j}$) with $j=1,2,3$ as $\theta \to \theta + \pi/2$($\theta_{23} \to \theta_{23} + \pi/2$) for 2RHN(3RHN) case. Thereupon, it suffices to anatomise the decays of only the first(first two) generation(s) of triplets for 2RHN(3RHN) case.

\section{\label{sec:dis}Displaced decays}
\noindent Dependence of the decay widths on the free parametres of the model, namely $x's$, $y$'s and the lightest neutrino mass, suggests that it is quite plausible that the heavy fermions of this model are long-lived for a certain region in the parametre space, and if so, disappearing track signature or other displaced vertex signals\footnote{The disappearing track signature arsing from $\Sigma^\pm$ decaying to $\Sigma^0$ and $\pi^\pm$ can be searched at the LHC if $\Sigma^0$ is long-lived enough to pass through the detector, {\it i.e.} $c\tau^0 \gtrsim \mathcal{O}(10)$ m, whereas such a displaced vertex signature can be observed at $ep$-colliders like FCC-he and LHeC if $c\tau^\pm \in [10^{-5},10^{-1}]$ m. Meanwhile, $\Sigma^0$ can be observed by MATHUSLA if $c\tau^0 > \mathcal{O}(100)$ m.} arising from them can be probed at collider. Ref.~\cite{sudip} and Ref.~\cite{Arindam_Sanjoy} discussed this possibility in detail. Here, we briefly mention our findings in this regard.


For 2RHN case, the decay lengths\footnote{For $M_i \gg M_{W,Z,h}$, the total decay lengths of $\Sigma^{0,\pm}_i$ are given by $c\tau^{0,\pm}_i \approx \times 1.975 \times 10^{-16} \times \left(\Gamma_{i,\rm T}^{0,\pm}\right)^{-1}~ {\rm GeV~m}$.} can not be arbitrarily large, withal the decay lengths can be as large as $\mathcal{O}(10^{-4})$ m when $y \sim 0$, see Eq.~\eqref{eq:2rhn_dl} and Eq.~\eqref{eq:2rhn_dl2}. While one doesn't expect any disappearing track signature arising from $\Sigma^\pm_i$ at the LHC, the same can be observed at FCC-he and LHeC.

For 3RHN case, the decay length for the neutral heavy fermions can be arbitrarily large as can be reckoned from Eq.~\eqref{eq:3rhn_dl}, whereas that of the charged ones could be as large as $\sim 5.6$ cm, and this has important implications of probing this scenario in various colliders. We briefly summarise the results for the first generation triplet fermion for NH in the following.
\begin{enumerate}
\item For $m_1 \lesssim \mathcal{O}(10^{-8})$ eV, $x_{2,3} \sim 0,\pi$ and $y_{2,3} \sim 0$, one gets $c\tau^0 \gtrsim \mathcal{O}(100)$ m and $c\tau^+ \sim 5.6$ cm. This is ideal for the MATHUSLA detector.
\item For $m_1 \lesssim \mathcal{O}(10^{-7})$ eV, $x_{2,3} \sim 0,\pi$ and $y_{2,3} \sim 0$, $c\tau^0 \gtrsim \mathcal{O}(10)$ m and $c\tau^+ \sim 5.6$ cm. This is ideal for the disappearing track signature searches at the LHC.
\item For $m_1 \lesssim \mathcal{O}(10^{-1})$ eV, $y_2 \lesssim \mathcal{O}(1)$ and $y_3 \lesssim \mathcal{O}(2)$, $c\tau^+$ lies in the range $[10^{-5},10^{-1}]$ m for all values of $x_2$ except $x_2 \sim \pi/2,3\pi/2$. For $x_2 \sim \pi/2,3\pi/2$, one would expect $c\tau^+$ to lie in the same range for $y_2 \sim 0$ and $y_3 \lesssim \mathcal{O}(40)$. This is ideal for the displaced vertex signature searches at FCC-he and LHeC.
\end{enumerate}
We see that for a certain region in the parametre space of $x_{1,2,3},y_{1,2,3}$ and $m_1$, the first generation of triplet fermion could be long-lived and, possibly, can be observed by the detectors. Likewise, the second (and hence third in view of the symmetry stated at the end of Section~\ref{sec:model}) generation of triplet fermion could also be long-lived for a certain region in the parametre space of $x_{1,2,3},y_{1,2,3}$ and $m_1$\footnote{Note that for the most trivial choice $R=\mathbb{1}$ ({\it i.e.} all $x,y$'s $\sim 0$), only $\Sigma^0_1(\Sigma^0_3)$ has arbitrarily large decay length as $m_1(m_3) \to 0$ for NH(IH); whereas for a non-trivial choice of $R$, $\Sigma^0_{2,3}(\Sigma^0_{1,2})$ could also have arbitrarily large decay length as $m_1(m_3) \to 0$ for NH(IH). For example, a choice like $y_{1,3} \sim 0$, $x_1 \sim 0[\pi/2]$ and $x_3 \to \pi/2,3\pi/2$ leads to arbitrarily large decay length for $\Sigma^0_2[\Sigma^0_3]$ as $m_1 \to 0$ for NH; whereas a choice like $x_2 \sim \pi/2,3\pi/2$ and $y_2 \sim 0$ leads to arbitrarily large decay length for $\Sigma^0_1$ as $m_3 \to 0$ for IH.}. A similar discussion follows for IH case also.

Though the quantity $\tilde{m}_i=\sum_j m_j \left|R_{ij}\right|^2$ can be inferred, as argued in Ref.~\cite{Hambye}, by measuring the displaced vertices if the triplet fermions are long-lived, the lightest neutrino mass can not be estimated unless the complex orthogonal matrix, $R$, is known beforehand. Since we do not know any viable way to measure the complex angles in $R$ yet, we can't infer the lightest neutrino mass even if we could measure the decay lengths. In fact, to do so, we have to know the $R$ matrix very precisely, specifically the values of the $y$'s. This is because the decay lengths have exponential dependence on the $y's$. Thus, even for hardly different values of $y$, the same $c\tau$ value may corresponds to completely different values of the lightest neutrino mass. However, one can extract some information regarding the complex angles in $R$ by measuring the decay lengths at the MATHUSLA or the LHC. For example, a positive search result at the MATHUSLA detector or the LHC would infer that $x_1\sim 0,\pi/2$, $x_2\sim 0,\pi$, $x_3\sim 0,\pi/2,\pi,3\pi/2$ and $y_{1,2,3}\sim 0$.

\section{\label{sec:multilep}Multilepton final states search}
Production of the triplet fermions in pair and their subsequent decays into SM two body final states lead to a variety of final state signatures including the multilepton final states signatures at the LHC. The CMS collaboration has recently published a multilepton final states search \cite{cms_137} with an integrated luminosity of $137.1$ fb$^{-1}$ of pp collisions at $\sqrt{s}=13$ TeV. Taking into account all possible multilepton final states in the type-III seesaw model with one generation of triplet fermions, the CMS analysis excluded them for masses below 880 GeV at $95\%$ confidence level (CL) in the flavour democratic scenario. Note that a realistic type-III seesaw scenario requires at least two generations of triplets; also, their decays are not necessarily flavour democratic. This is why the CMS observed 95$\%$ CL upper limits \cite{cms_137} on the total pair production cross-sections of triplet fermions are not straightforwardly applicable for such a realistic type-III seesaw model. To this end, we extend this search in the context of a realistic type-III seesaw model with two or three generations of triplet fermions, and set forth to explore the collider implications of the arbitrary complex orthogonal matrix, $R$. Thereupon, we restrict ourselves in the case where the heavy fermions produced at the LHC decay promptly inside the detector, and to ensure that we assume $m_1(m_3) > 10^{-3}$ eV for NH(IH) in 3RHN case.

We have implemented the model in SARAH \cite{sarah,sarah2} to generate UFO modules. The signal events are simulated using MadGraph \cite{mg5} with the \textit{NNPDF23LO} \cite{nnpdf,nnpdf2} parton distribution function, and showered with PYTHIA \cite{pythia}, and then DELPHES \cite{delphes} is used for detector simulation. Since the relevant SM backgrounds in our analysis are exactly same as that of Ref.~\cite{cms_137}, we do not simulate the backgrounds, instead we use distributions of expected SM backgrounds and observed events given in figure 3 and 4 of Ref.~\cite{cms_137} and also in \cite{cms137hepdata}. We closely follow the multilepton search strategies of Ref.~\cite{cms_137} for object reconstruction and selection, defining signal regions  and event selection. This search considers only the final states with three or more leptons ($e,\mu$). Based on the number of leptons, number of opposite-sign same-flavour (OSSF) lepton pairs and the invariant mass of OSSF pair, $\rm{M_{OSSF}}$, the events are categorised into seven signal regions (SRs), namely {\it 4LOSSF0, 4LOSSF1, 4LOSSF2, 3LOSSF0, 3L below-Z, 3L on-Z} and {\it 3L above-Z}\footnote{The names of the 4 (4L) or more lepton signal regions are self explanatory. The events with exactly three leptons containing an OSSF lepton pair are categorised based on $\rm{M_{OSSF}}$, and they are labelled by {\it 3L below-Z, 3L on-Z} and {\it 3L above-Z} when $\rm{M_{OSSF}}$ is below, within and above the Z-boson mass window ($\rm{M_Z \pm 15}$), respectively. The detailed description of object reconstructions, event selections and signal region definitions can be found in Ref.~\cite{cms_137}. Whereas, our implementation and validation of the CMS multilepton search strategies in Ref.~\cite{cms_137} were presented in Ref.~\cite{project1}.}. The signal regions are further classified into 40 statistically independent signal bins using $\rm{L_T + p_T^{miss}}$ or the transverse mass\footnote{The transverse mass is defined as $M_T=\sqrt{2 p_T^{\rm miss} p_T^\ell [1-\cos(\Delta\phi_{\vec p_T^{\rm miss},\vec p_T^\ell})]}~,$ where $\vec p_T^\ell$ is the transverse momentum vector of the lepton which is not a part of the on-Z pair, and $\Delta\phi_{\vec p_T^{\rm miss},\vec p_T^\ell}$ is the azimuthal separation between $\vec p_T^{\rm miss}$ and $\vec p_T^\ell$.} as the primary kinematic discriminant where $\rm{L_T}$ is the scalar sum of transverse momentum of all charged leptons, $\rm{ p_T^{miss}}$ is the missing transverse momentum. We use a hypothesis tester named `Profile Likelihood Number Counting Combination' \cite{bayes} to estimate CL. See \cite{project1} for details.

\begin{figure}[htb!]
\centering
\includegraphics[scale=0.6]{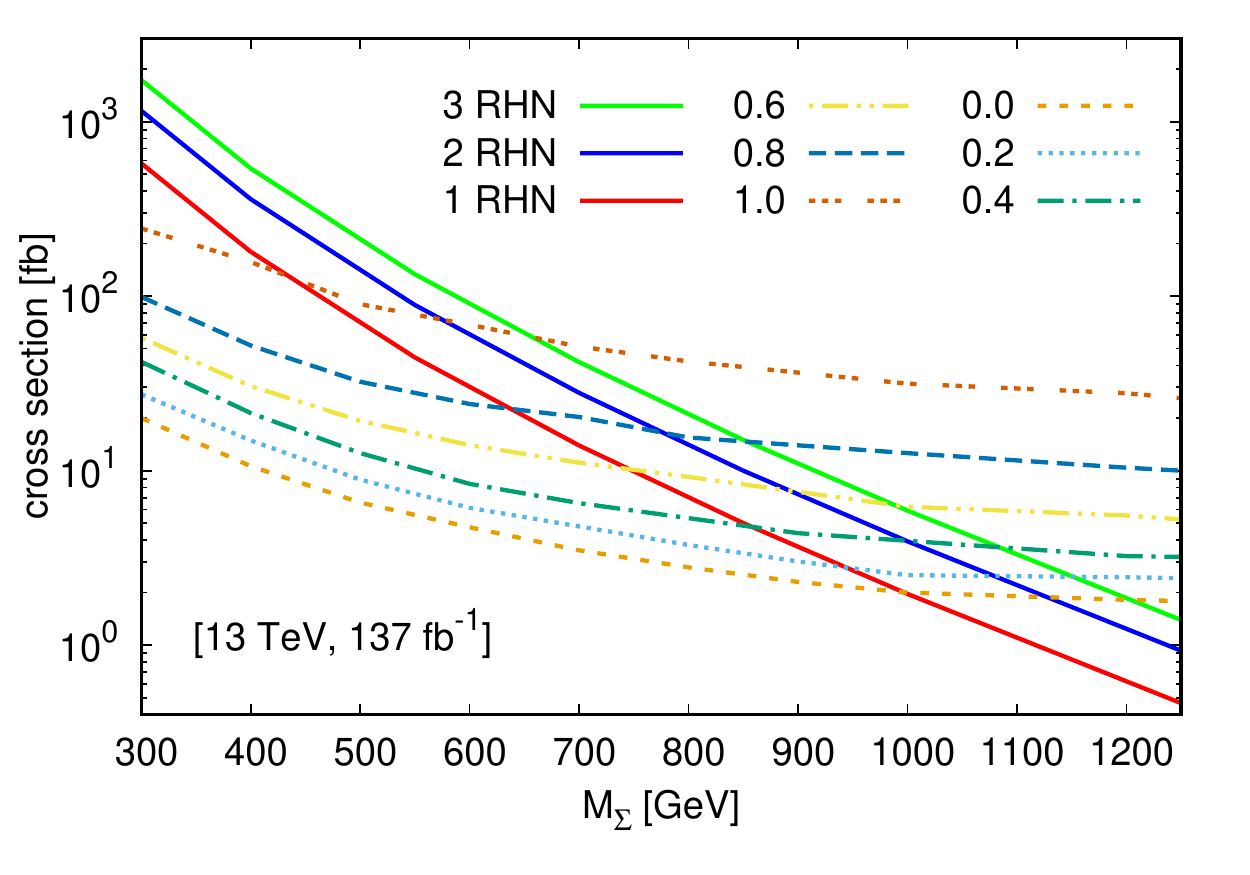}
\caption{$95\%$ CL upper limit on total production cross section of heavy fermions for different values of ${\rm BR}_\tau$. Solid red line shows theoretical prediction for the total cross section at LO+NLO for one generation of triplet fermions. }
\label{fig:CL}
\end{figure}

\begin{figure}[htb]
\centering
\includegraphics[scale=0.6]{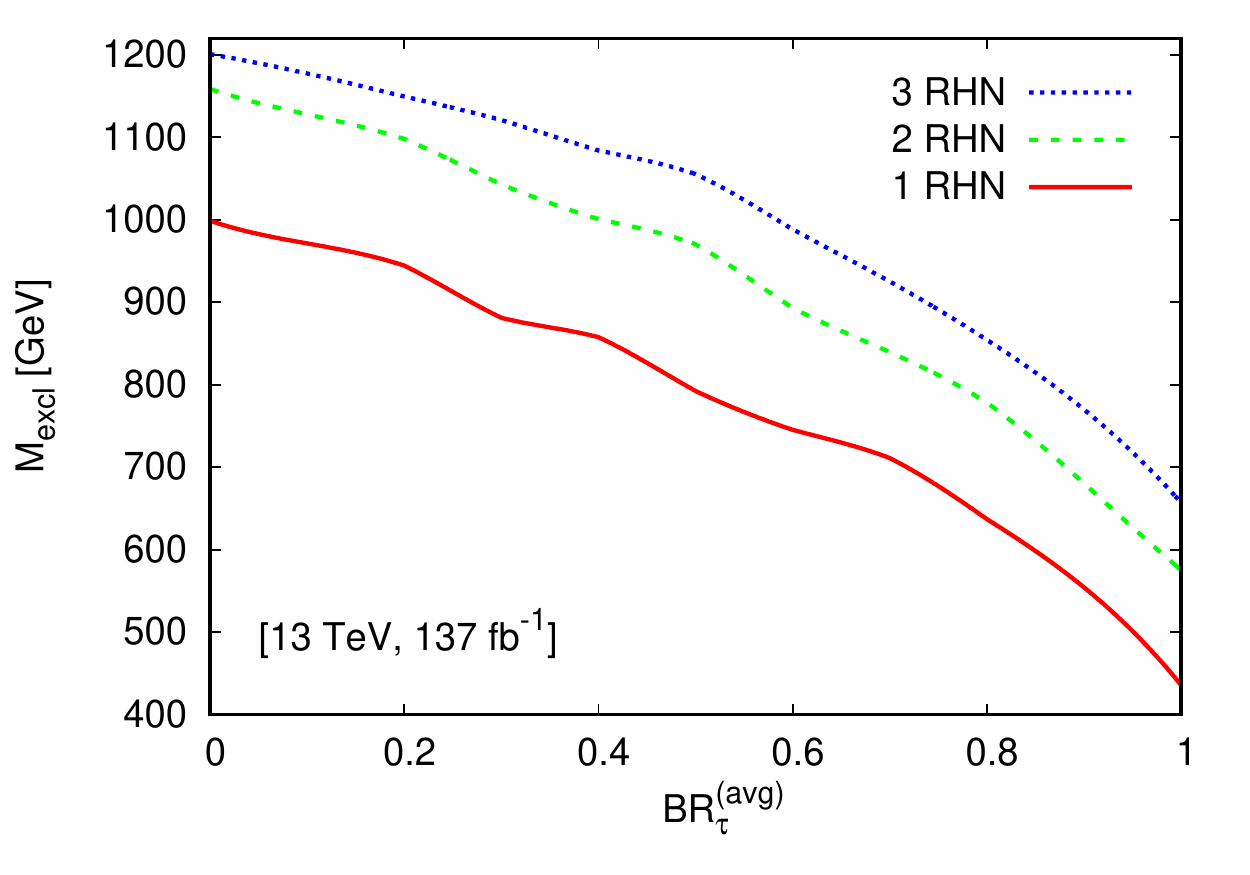}
\caption{$95\%$ CL lower limit on the degenerate mass of the triplet fermions as a function of $\rm{BR_\tau^{\rm avg}}$.}
\label{fig:M_excl}
\end{figure}

For convenience, we define total branching ratios of neutral and charged component of the $i^{\rm th}$ generation triplet fermion ($\Sigma^0_i$ and $\Sigma^\pm_i$) to the lepton flavour $\ell_\alpha (=e,\mu,\tau)$ as: $\mathrm{BR}^0_{i,\alpha}=\mathrm{BR}\left(\Sigma^0_i \to \ell_\alpha^\pm W^\mp\right)$ and $\mathrm{BR}^\pm_{i,\alpha}=\mathrm{BR}\left(\Sigma^\pm_i \to \ell_\alpha^\pm Z\right)+\mathrm{BR}\left(\Sigma^\pm_i \to \ell_\alpha^\pm h\right)$, subject to $\sum_\alpha {\rm BR}^{0,\pm}_{i,\alpha}=1$. For $M_i \gg M_{W,Z,h}$, $\mathrm{BR}^0_{i,\alpha} \approx \mathrm{BR}^\pm_{i,\alpha}$ (see Eq.~\eqref{eq:DW_Sigma_XY}); henceforward, we will not distinguish between $\mathrm{BR}^0_{i,\alpha} \mathrm{and~BR}^\pm_{i,\alpha}$. Moreover, it is not possible to distinguish among the degenerate (in mass) copies of triplets at the LHC unless a big conspiracy makes these copies very different from one another. Therefore, in case of degenerate copies, instead of branching ratios for individual copies, we use the average branching ratio defined as:
\begin{equation}
\label{eq:BR_avg}
\mathrm{BR}_\alpha^{\rm avg} =\frac{1}{N} \sum_{i=1}^N \mathrm{BR}_{i,\alpha}=\frac{1}{N} \sum_{i=1}^N \frac{|(Y_\Sigma)_{i\alpha}|^2}{\sum_\alpha |(Y_\Sigma)_{i\alpha}|^2}~,
\end{equation}
where $N=2(3)$ for 2RHN(3RHN) case. Noting that CMS has high reconstruction and identification efficiency for both electrons and muons, for the sake of simplicity, we assume equal reconstruction and identification efficiency for both electrons and muons. This assumption does not affect the estimated limits significantly\footnote{For example, for ${\rm BR}_e={\rm BR}_\mu=50\%$, the $95\%$ CL lower limit on the triplet mass is 1000 GeV (see Figure~\ref{fig:CL} and Figure~\ref{fig:M_excl}); if one considers ${\rm BR}_e=100\%$ (${\rm BR}_\mu=100\%$), the limit would be relaxed(improved) by not more than 20(10) GeV which is well within a few percent level.}. Therefore, enduring the margin of a few percent, the $95\%$ CL limits doesn't depend on ${\rm BR}_e$ and ${\rm BR}_\mu$ exclusively but on ${\rm BR}_e+{\rm BR}_\mu=1-{\rm BR}_\tau$.

Bearing the aforesaid discussions in mind, we estimate $95\%$ CL upper limit on the total production cross section of triplet fermionic pairs for different branching ratio to the $\tau$-lepton final state (see Figure~\ref{fig:CL}) following the approach prescribed in Ref.~\cite{project1}. The theoretical predictions for the total triplet production cross section at LO+NLO\footnote{The cross sections at LO+NLO are taken from Ref.~\cite{cms_137}.} are also presented in Figure~\ref{fig:CL}.
Figure~\ref{fig:M_excl} shows $95\%$ CL lower limit on the (degenerate) mass of the triplet fermions, $\rm{M_{excl}}$, as a function of (average) branching ratio to the $\tau$-lepton final state for one, two and three degenerate copies of triplet fermions. Since the CMS analysis \cite{cms_137} is dedicated to search for multi-electron/muon final states, the most restrictive bound on $\rm{M_{excl}}$ is found for $\tau$-phobic scenario, {\it i.e.} $\rm{BR_\tau^{\rm (avg)}}=0$, whereas the least restrictive bound corresponds to $\tau$-philic scenario, {\it i.e.} $\rm{BR_\tau^{\rm (avg)}}=1$.

The remainder of this letter is devoted to comprehend the consequence of the free parametres in $R$ on the $95\%$ CL lower limit on the triplet masses in a realistic type-III seesaw with two or three generations of triplet fermions. As stated at the closing of Section~\ref{sec:model} and Section~\ref{sec:pro_dec}, because of the symmetry $R_{1j} \to R_{2j}$ ($R_{2j} \to R_{3j}$) with $j=1,2,3$ as $\theta \to \theta + \pi/2$ ($\theta_{23} \to \theta_{23} + \pi/2$) for 2RHN (3RHN) case, it suffices to anatomise the decays of only the first (first two) generation(s) of triplets for 2RHN (3RHN) case. Noting this, we explicitly study only the first (first two) generation(s) of triplets for 2RHN (3RHN) case.

\subsection{2RHN case}
\begin{figure*}[htb!]
\centering
\includegraphics[scale=0.133]{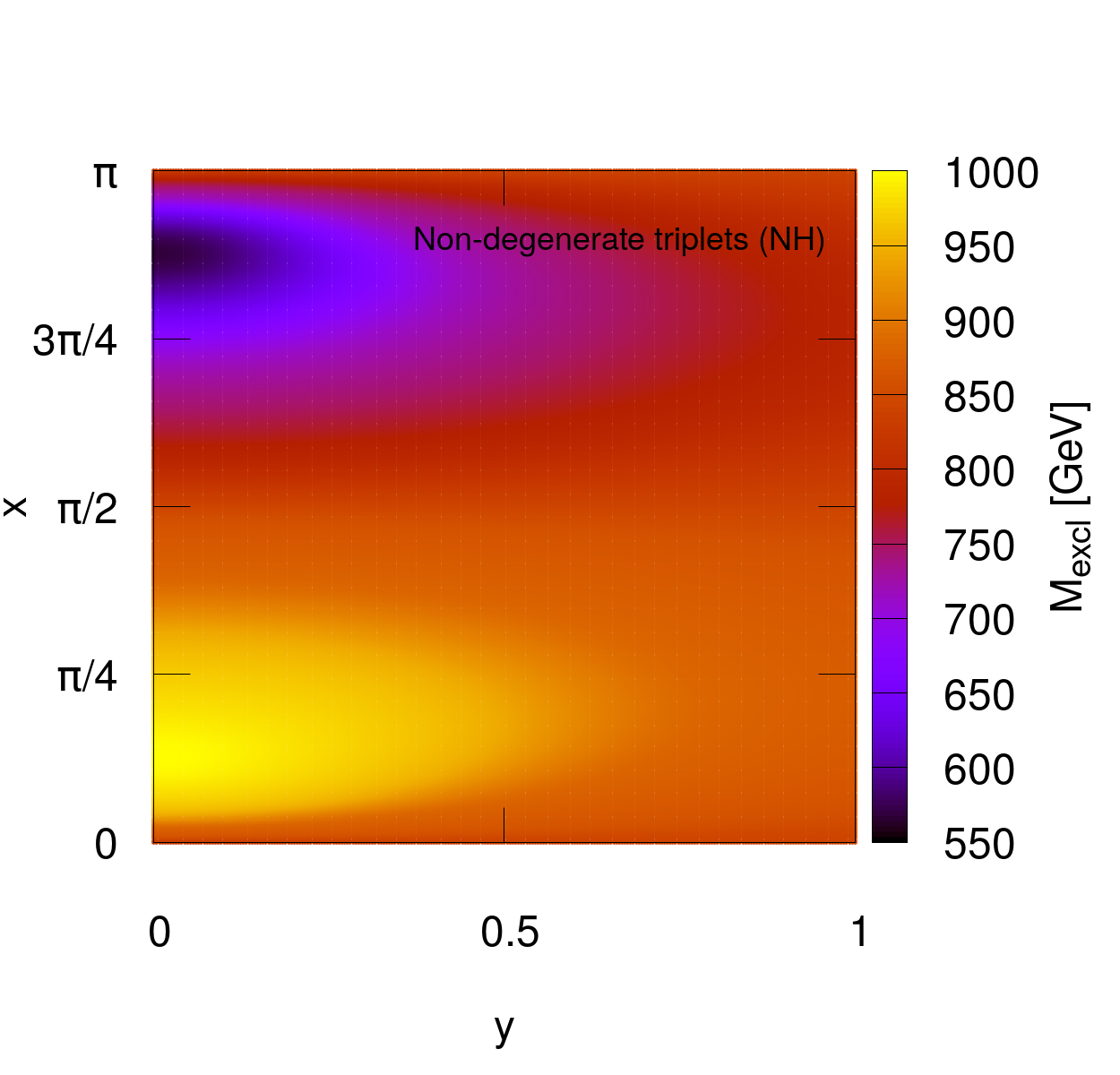}
\includegraphics[scale=0.133]{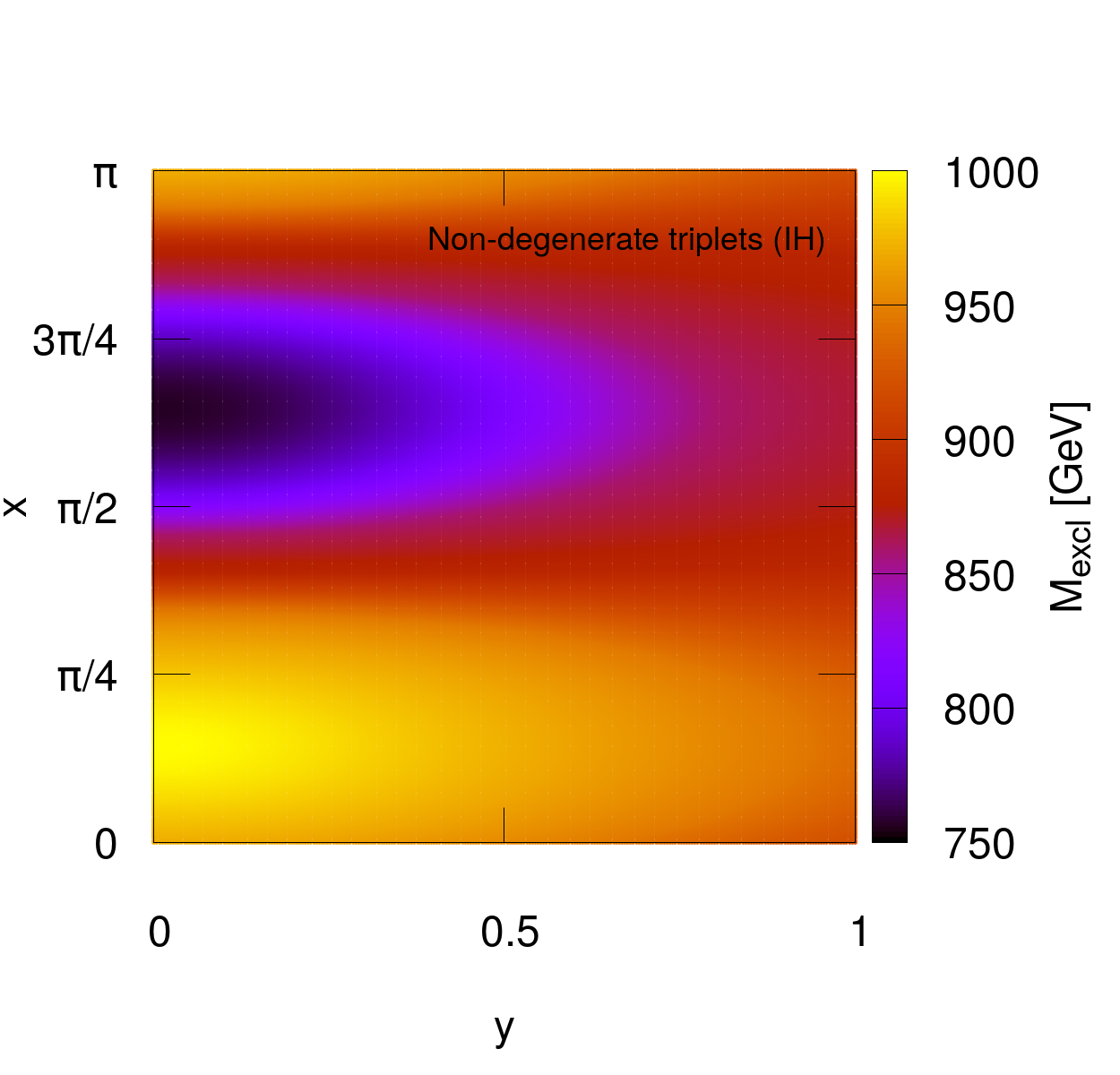}
\includegraphics[scale=0.133]{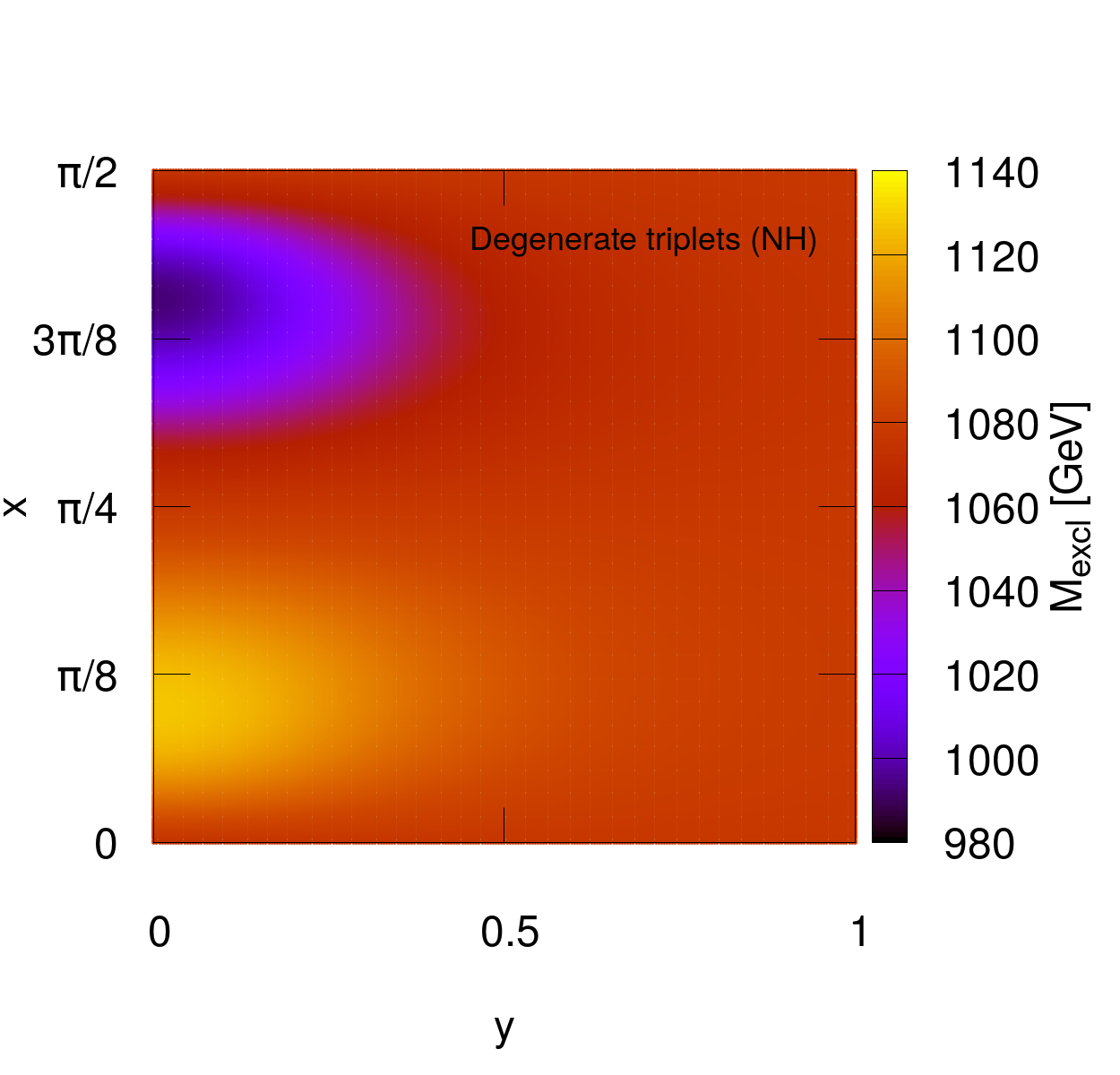}
\caption{To the left(middle): $95\%$ CL lower limit on the mass, shown by the colour gradient, of the first generation of triplet fermions in GeV in the $x$-$y$ plane for NH(IH). To the right: $95\%$ CL lower limit on the degenerate mass of the triplet fermions in GeV in the $x$-$y$ plane for NH.}
\label{fig:2RHN_1_Mexcl}
\end{figure*}

\noindent For 2RHN case, there are two real free parametres in $R$ --- $x (\in [0,\pi]) ~\mathrm{and}~ y (\geq 0)$. Left(middle) panel in Figure~\ref{fig:2RHN_1_Mexcl} shows $95\%$ CL lower limit on the mass ($\rm{M_{excl}}$) of the first generation of triplet fermions in the $x$-$y$\footnote{For large enough $y$ ($y>\mathcal{O}(1)$), $|\cos\theta| \approx |\sin\theta| \approx e^y/2$ and thus, the normalised branching ratios and hence, the resulting collider bound, $M_{\rm excl}$, become independent of both $x$ and $y$ --- $\rm BR_e \approx 6\%(49\%)$, $\rm BR_\mu \approx 50\%(22\%)$ and  $\rm BR_\tau \approx 43\%(29\%)$ for NH(IH).} plane for NH(IH) assuming that the second generation of the same are too heavy to be produced so that effectively only the first generation fermions are produced at the LHC. The bound on $\rm{M_{excl}}$ ranges over 565--1000 (775--1000) GeV for NH(IH). For NH, the most(least) restricted bound is found for $x \sim 23^\circ$ and $y \sim 0$ ($x \sim 158^\circ$ and $y \sim 0$) which corresponds to ${\rm BR}_\tau \sim 89\%(0\%)$. Whereas, for IH, the most(least) restricted bound is found for $x \sim 26^\circ$ and $y \sim 0$ ($x \sim 116^\circ$ and $y \sim 0$) which corresponds to ${\rm BR}_\tau \sim 58\%(0\%)$. For the second generation of triplet fermions, the bounds on $\rm{M_{excl}}$ are exactly same as that of the first generation ones in Figure~\ref{fig:2RHN_1_Mexcl} with the $x$-coordinates replaced by $x+\pi/2$\footnote{This is a consequence of the symmetry $R_{1j} \to R_{2j}$ ($j=1,2,3$) as $\theta \to \theta + \pi/2$.}.

Right panel in Figure \ref{fig:2RHN_1_Mexcl} shows $95\%$ CL lower limit on the degenerate mass of triplet fermions for NH. The bound on $\rm{M_{excl}}$ ranges over 990--1130 GeV. The most(least) restricted bound is found for $x \sim 18^\circ$ and $y \sim 0$ ($x \sim 72^\circ$ and $y \sim 0$) which corresponds to ${\rm BR}_\tau \sim 60\%(28\%)$. For IH, the corresponding lower limit is $\sim 1050$ GeV for all values of $x$ and $y$\footnote{For IH, the first and second generation of light neutrinos becomes almost degenerate, {\it i.e.} $m_1 \approx m_2$, and Eq.~\eqref{eq:BR_avg} reduces to $\mathrm{BR}_\alpha^{\rm avg} \approx \frac{1}{2} (1-|U_{\alpha 3}|^2)$. Thus, the branching ratios become independent of both $x$ and $y$ and so is the $95\%$ CL limit.}. 


\subsection{3RHN case}
\begin{figure*}[htb!]
\centering
\includegraphics[scale=0.133]{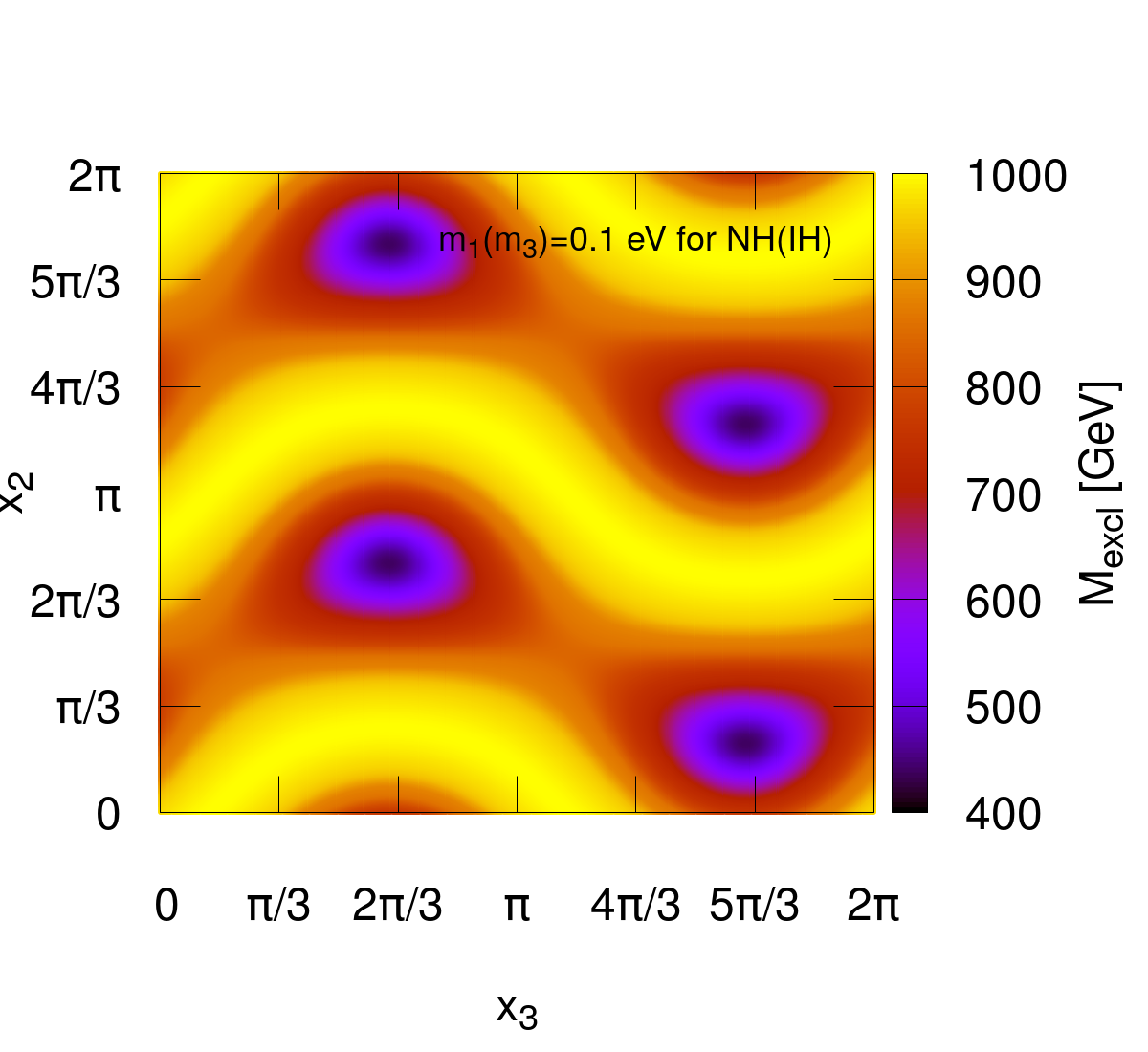}
\includegraphics[scale=0.133]{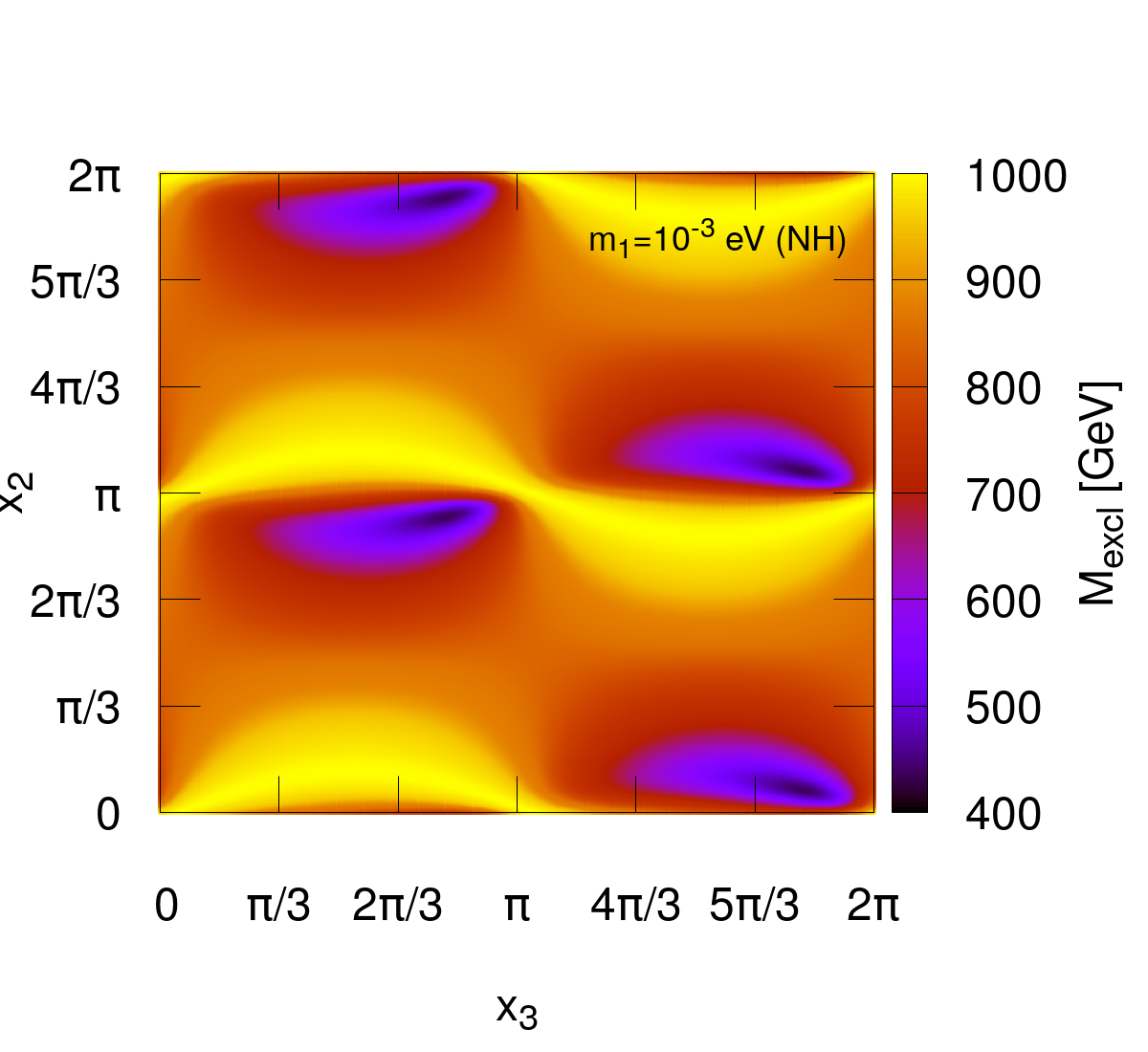}
\includegraphics[scale=0.133]{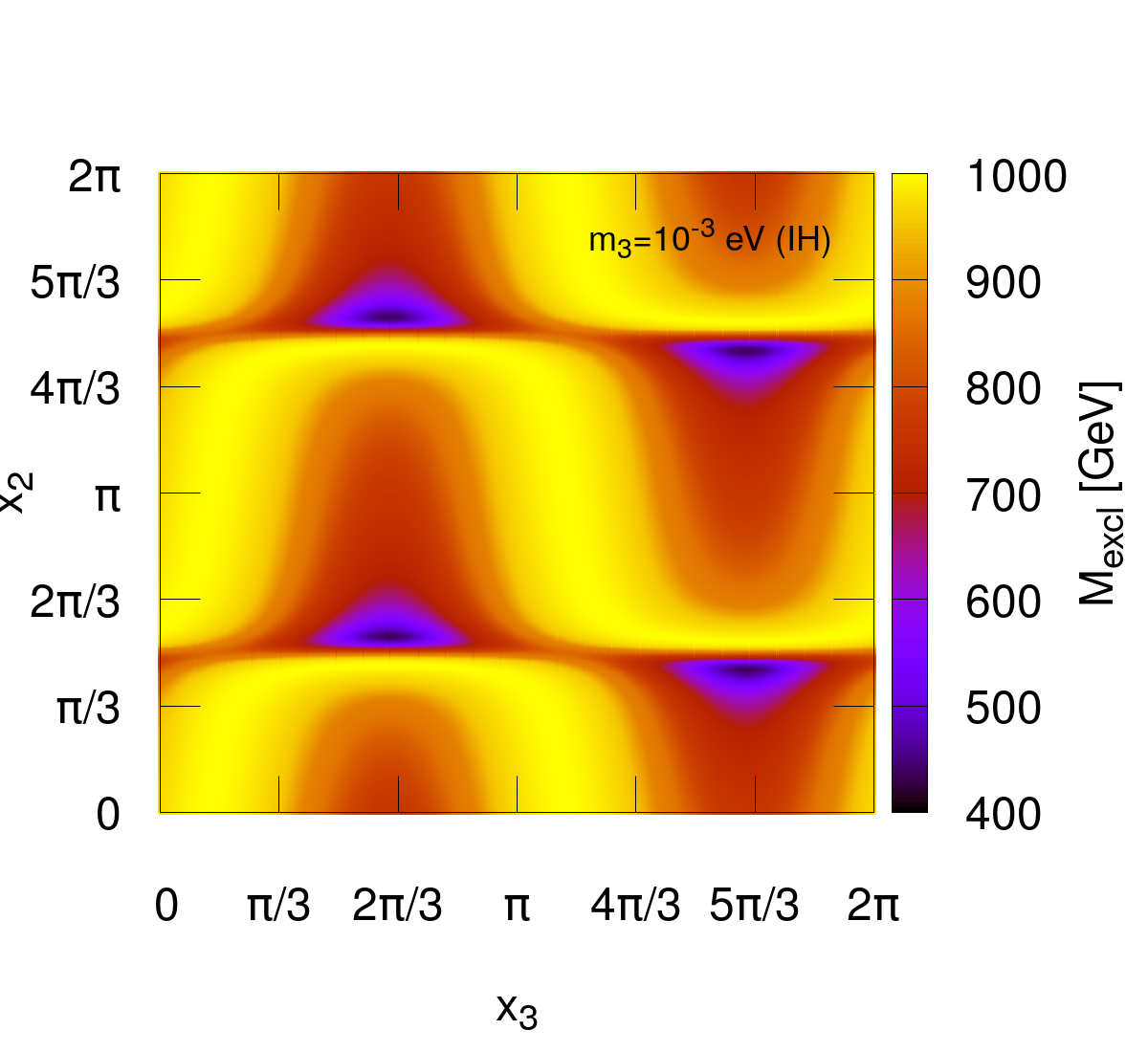}
\caption{$95\%$ CL lower limit on the mass of the first generation of triplet fermions in GeV in the $x_2$-$x_3$ plane. The left panel corresponds to $m_1(m_3)=0.1$ eV for both NH and IH. The middle(right) panel correspond to $m_1(m_3)=10^{-3}$ eV for NH(IH).}
\label{fig:3RHN_Re_1_Mexcl}
\end{figure*}

\begin{figure*}[htb!]
\centering
\includegraphics[scale=0.35]{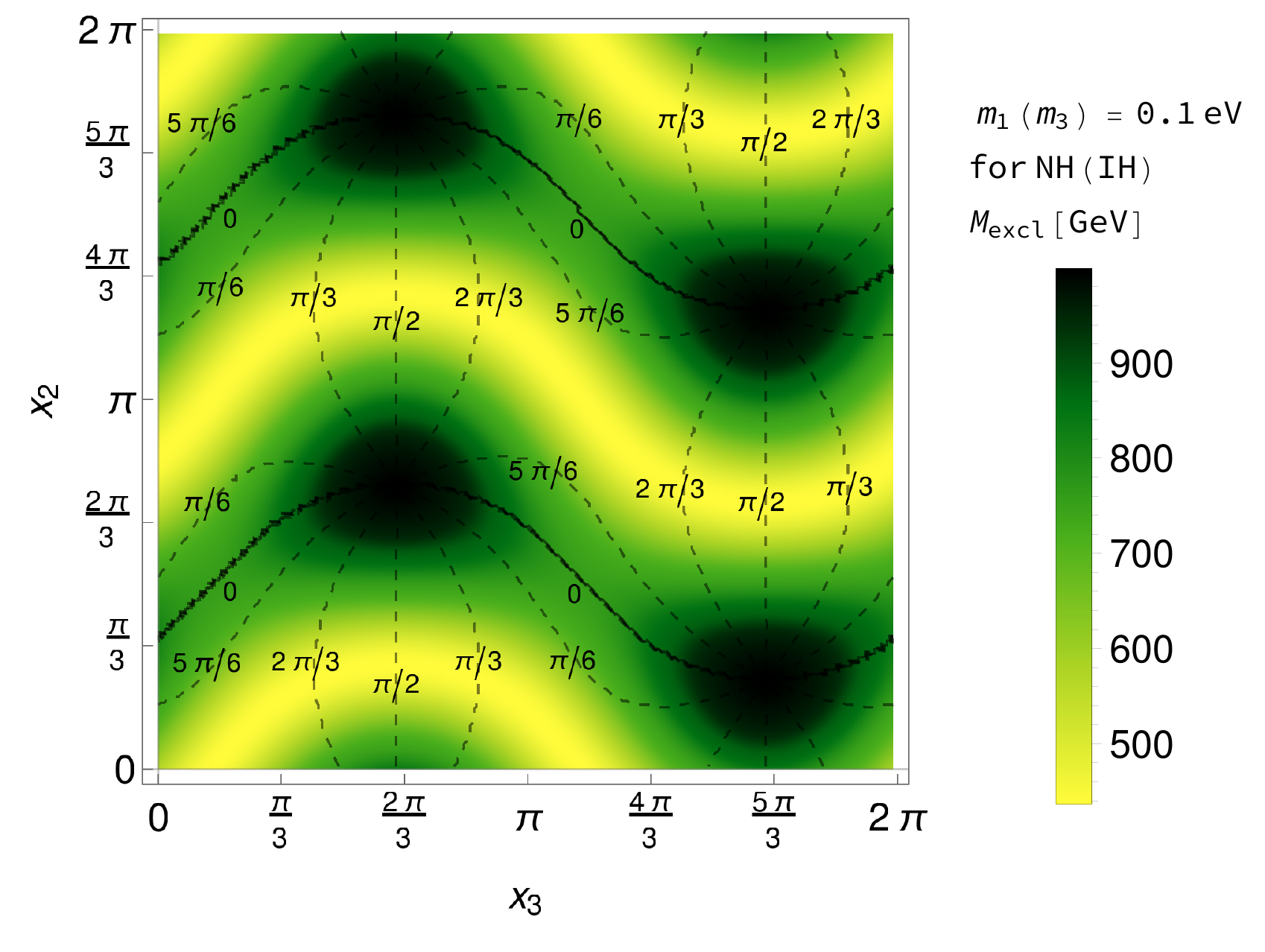}
\includegraphics[scale=0.35]{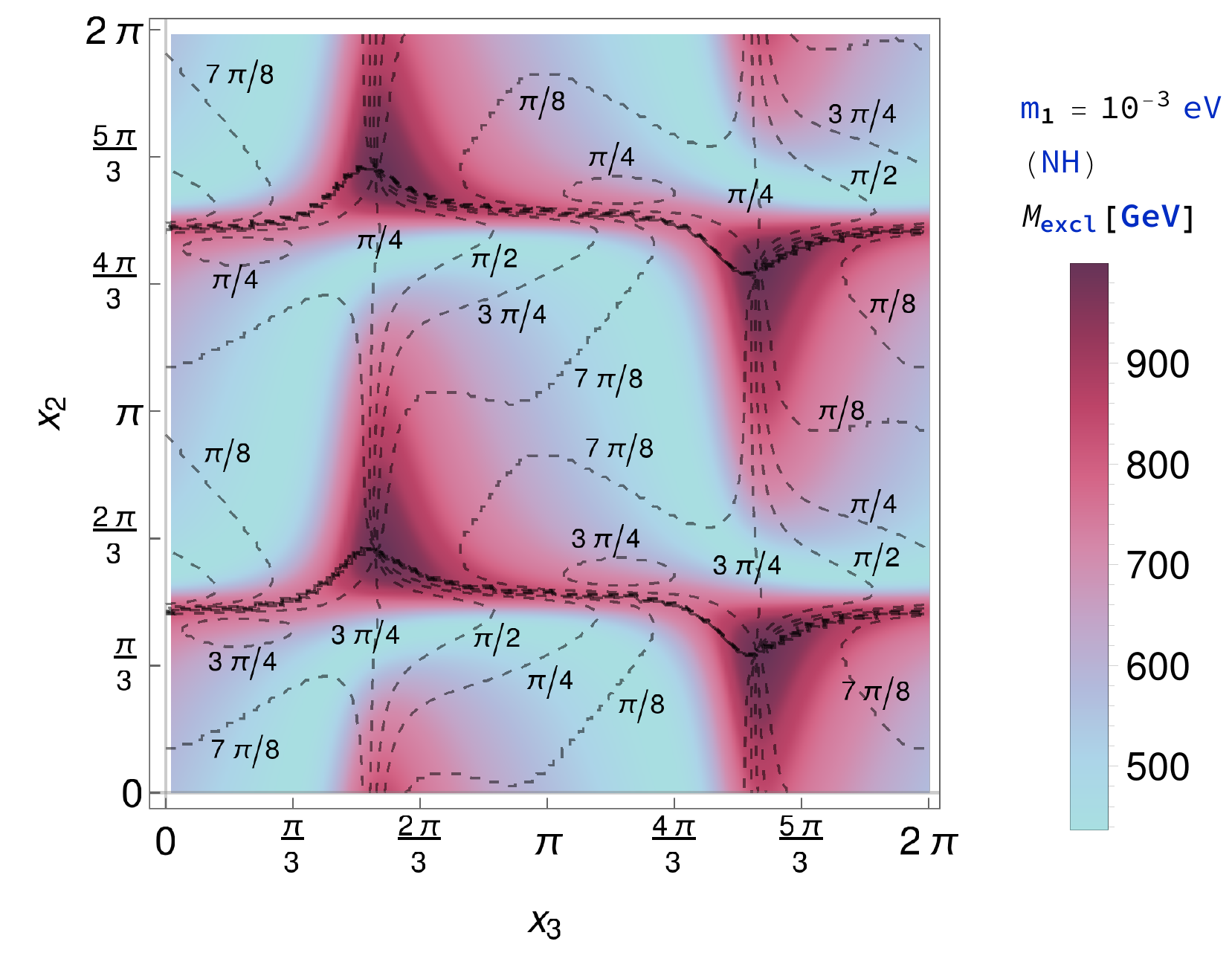}
\includegraphics[scale=0.35]{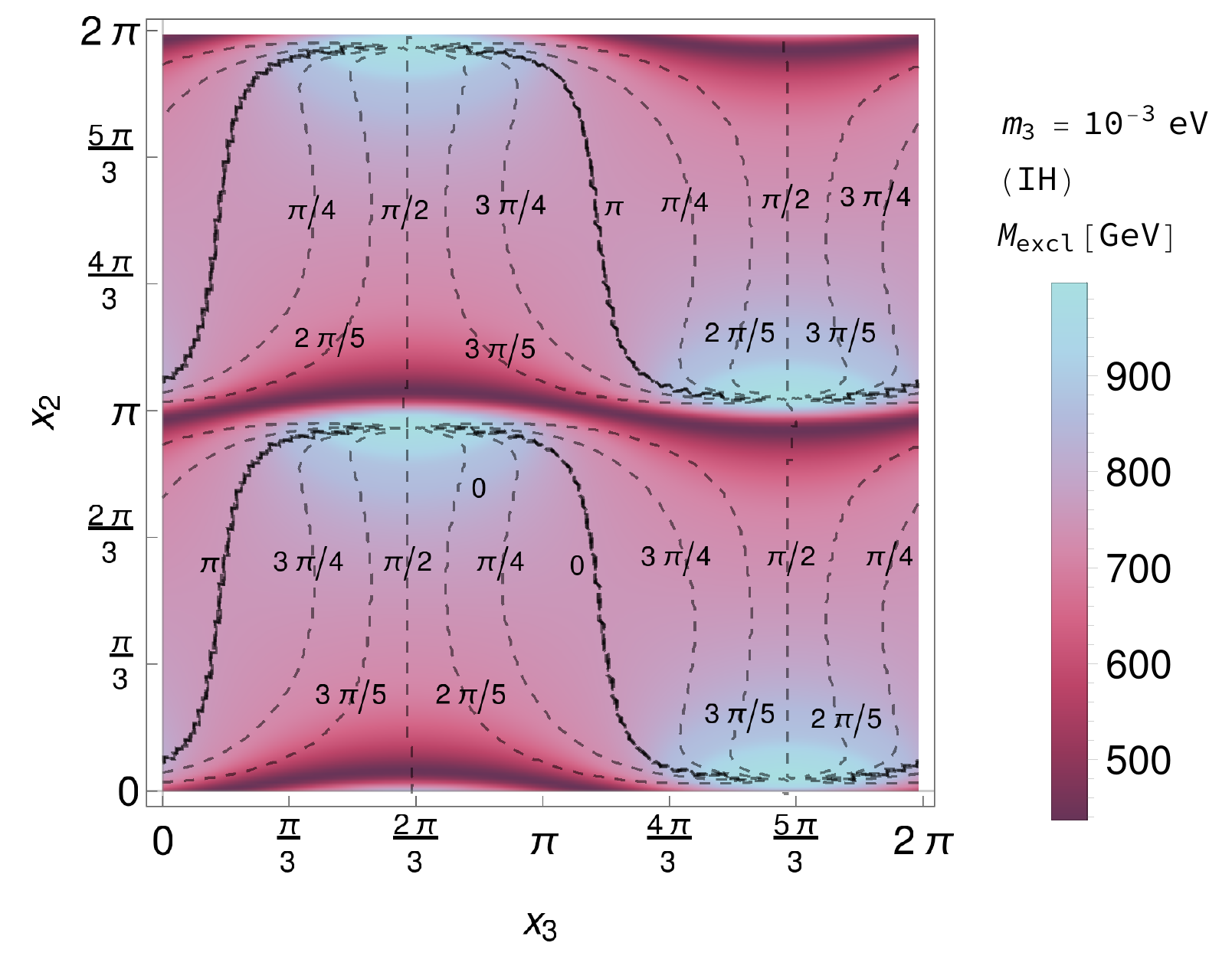}
\caption{Least restrictive $95\%$ CL lower limit on the mass of the second generation of triplet fermions in GeV along with the contours for $x_1$ in the $x_2$-$x_3$ plane. The left panel corresponds to $m_1(m_3)=0.1$ eV for both NH and IH. The middle(right) panel correspond to $m_1(m_3)=10^{-3}$ eV for NH(IH).}
\label{fig:3RHN_Re_2_Mexcl}
\end{figure*}

\noindent For 3RHN case, there are six real free parametres in $R$ --- $x_1 (\in [0,\pi]), x_{2,3} (\in [0,2\pi]) ~\mathrm{and}~ y_{1,2,3}$. For the sake of simplicity, we assume that $R$ is real, {\it i.e.} $y_{1,2,3} \sim 0$\footnote{Implications of large $y$'s are discussed at the end of this section.}. The normalised branching ratios for the first generation of triplet fermions are independent of $x_1$ (see Eq.~\eqref{eq:casas_ibarra} and \eqref{eq:DW_Sigma_XY}), and so are the $95\%$ CL lower limits on the mass. Left panel in Figure~\ref{fig:3RHN_Re_1_Mexcl} shows $95\%$ CL lower limit on the mass of the first generation of triplet fermions in GeV in the $x_2$-$x_3$ plane for $m_1(m_3)=0.1$ eV for both NH and IH\footnote{Note that for the lightest neutrino mass of $\mathcal{O}(0.1)$ eV, the light neutrinos become almost degenerate in mass making the hierarchy irrelevant. This has been reflected in Figure~\ref{fig:3RHN_Re_1_Mexcl}(left panel) and also in Figure~\ref{fig:3RHN_Re_2_Mexcl}(left panel).} assuming that effectively only the first generation of triplet fermions are produced at the LHC. Likewise, middle(right) panel in Figure~\ref{fig:3RHN_Re_1_Mexcl} shows the same for $m_1(m_3)=10^{-3}$ eV for NH(IH). In all these cases, depending on the values of $x_2$ and $x_3$, the limits range over 435--1000 GeV.

Unlike the first generation, the normalised branching ratios for the second generation of triplet fermions, and hence the $95\%$ CL lower limit on the mass of the same depend on all the three parametres --- $x_{1,2,3}$. So, in the second generation case, we estimate least restrictive $95\%$ CL lower limit on the mass in the $x_2$-$x_3$ plane. To rephrase it, for a given set of $x_2$ and $x_3$, we vary $x_1$ to find the most conservative limit. The right panel in Figure~\ref{fig:3RHN_Re_2_Mexcl} shows least restrictive $95\%$ CL lower limit on the mass of the second generation of triplet fermions in the $x_2$-$x_3$ plane for $m_1(m_3)=0.1$ eV for NH(IH). Likewise, the middle(right) panel in Figure~\ref{fig:3RHN_Re_2_Mexcl} shows the same for $m_1(m_3)=10^{-3}$ eV for NH(IH). In Figure~\ref{fig:3RHN_Re_2_Mexcl}, the contours, in the $x_2$-$x_3$ plane, show the values of $x_1$ for which the corresponding limit on the mass is least restrictive. Figures~\ref{fig:3RHN_Re_2_Mexcl} and \ref{fig:3RHN_Re_1_Mexcl} clearly show that the type-III seesaw triplets with mass as low as 435 GeV is still allowed in certain parts of $x_1$--$x_2$--$x_3$ parameter space. Note that this limit is substantially smaller than the limit derived in Ref.~\cite{cms_137} in the context of simplified type-III seesaw.

Now, we consider the scenario when three degenerate copies of triplet fermions are produced at the LHC. For $m_1(m_3)\sim 0.1$ eV for NH(IH), {\it i.e.} for degenerate mass spectrum for the light neutrinos, the average branching ratio to different leptonic final states become flavour universal: Eq.~\ref{eq:BR_avg} simplifies to ${\rm BR}_\alpha^{\rm avg} \approx 1/3$ with $\alpha=e,\mu,\tau$. In such so-called flavour democratic scenario, the $95\%$ CL lower limit on the mass of the triplet fermions is found to be $\sim$ 1110 GeV. At this point, we accentuate that the flavour democratic scenario does not necessarily be an ad-hoc assumption, rather it could be a consequence of degenerate mass spectrum for both the light and heavy neutrinos \footnote{We briefly mention about two other special cases which also give rise to flavour democratic scenario:
(i) For the most trivial choice of $R$, {\it i.e.} $R=\mathbb{1}$, the branching ratio of $\Sigma^{0,\pm}_{i}$ to $\ell_\alpha$ becomes $|U_{\alpha i}|^2$. If one assumes the triplet fermions to be degenerate, then this gives rise to flavour democratic scenario. (ii) For degenerate mass spectrum of the light neutrinos, if one identifies the complex angles in $R$ as the mixing angles in the PMNS matrix ({\it i.e.}~$R=U$), the Yuakawa matrix ($Y_\Sigma$) becomes flavour diagonal. If one further assumes that the triplet fermions are degenerate, then this reduces to flavour democratic scenario.}.


For smaller values of the lightest neutrino mass, flavour universality of the the average branching ratio of the triplets to different leptonic final states breaks down. For brevity, we do not show any plots for this case. However, we note that for $m_1(m_3)=10^{-3}$ eV in NH(IH), the limit on the degenerate triplet mass ranges over 935-1150 (1045-1150) GeV.


Lastly, we briefly mention the implications of large $y$'s. For large enough $y$'s, $|\cos\theta$'s$| \approx |\sin\theta$'s$|\approx \frac{1}{2} e^y$'s. Thus, the normalised branching ratios become independent of all $x$'s and $y$'s, and so are the limits. For NH(IH) with $m_1(m_3)=0.1$ eV, the limit is 890 GeV corresponding to $\rm BR_\tau \sim 29\%$ for both the first and second generation of triplets. For NH(IH) with $m_1(m_3)=10^{-3}$ eV, the limit is 845(890) GeV corresponding to $\rm BR_\tau \sim 42\%(29\%)$ for both the first and second generation of triplets. Likewise, in the case of three degenerate  copies of triplet fermions, the limit is 1125 GeV corresponding to $\rm BR_\tau \sim 28\%$ for NH(IH) with $m_1(m_3)=0.1$ eV, and for NH(IH) with $m_1(m_3)=10^{-3}$ eV, the limit is 1080(1125) GeV corresponding to $\rm BR_\tau \sim 42\%(28\%)$.

\section{Summary and Outlook}
\noindent The high-energy see-saw theory involves 15(9) effective parametres, whereas the low-energy neutrino phenomenology involves 9(7) physical and measurable parametres in 3RHN(2RHN) case. A number of parametres get lost in integrating the heavy fields out. The well-known Casas-Ibarra parametrisation facilitates to encode the information lost in the decoupling of the heavy fermions in an arbitrary complex orthogonal matrix ($R$) with 6(2) real parametres in 3RHN(2RHN) case. We have explored the phenomenological implications of the said matrix in view of lepton flavour violation, displaced decays and a recent multilepton final states search by the CMS collaboration. We have seen that for a certain region in the parametre space of $x_{1,2,3} , y_{1,2,3}$ , all generations of triplet fermions could be long-lived; and possibly, a limited region of that parametre space can be probed by the detectors like MATHUSLA, LHC or FCC-he and LHeC. This is in contrast to the most trivial choice of $R$ ($=\mathbb{1}$) in which only the first(third) generation of triplet fermions could be long-lived. We accentuate that the lightest neutrino mass can not be inferred by measuring the displaced vertices unless the orthognal matrix is known beforehand. However, one can extract some information regarding the complex angles in R by measuring the decay lengths at the MATHUSLA or the LHC. For example, a positive search result at the MATHUSLA detector or the LHC would infer that $x_1\sim 0,\pi/2$, $x_2\sim 0,\pi$, $x_3\sim 0,\pi/2,\pi,3\pi/2$ and $y_{1,2,3}\sim 0$. We have explored collider implications of the arbitrary complex orthogonal matrix using a recent multilepton final states search by the CMS collaboration. Specifically, we have explored the consequence of the free parametres in $R$ on the $95\%$ CL lower limit on the mass of the triplet fermions in view of the multilepton final states search for both 2RHN and 3RHN case. We note that degenerate mass spectrum for both the light and heavy neutrinos lead to a flavour democratic scenario willy-nilly. In such-scenario, the $95\%$ CL lower limit on the degenerate mass of triplet fermions is found to be $\sim$1110 GeV. Depending on the choice for the said matrix, we find that the bounds could be notably contrasting than that obtained by the CMS search.


\end{document}